\documentclass[%
  aps,
  prd,
  reprint,
  superscriptaddress,
  nofootinbib,
  longbibliography
]{revtex4-2}

\usepackage[T1]{fontenc}
\usepackage[utf8]{inputenc}

\usepackage{amsmath,amssymb,slashed,bm}
\usepackage{gauss}
\usepackage{ytableau}

\usepackage{graphicx}
\usepackage{float}
\usepackage{subfig}
\usepackage[justification=raggedright,singlelinecheck=false]{caption}
\graphicspath{{fig/}{img/}}

\usepackage[colorlinks=true,linkcolor=blue,citecolor=blue,urlcolor=blue]{hyperref}

\newcommand{\Hc}{\mathbb{H}}
\newcommand{\Ec}{\mathbb{E}}
\newcommand{\Ht}{\widetilde{\mathbb H}}

\newcommand{\dd}{\mathrm{d}}
\newcommand{\ii}{\mathrm{i}}

\newcommand{\artanh}{\operatorname{artanh}}

\newcommand{\bperp}{\bm b_\perp}
\newcommand{\Dperp}{\bm\Delta_\perp}

\begin{document}

\title{Rapidity\texorpdfstring{‑}{-}Dependent Spin Decomposition of the Nucleon}

\author{Florian Hechenberger}\email{florian.hechenberger@stonybrook.edu}
\affiliation{Center for Nuclear Theory, Stony Brook University, Stony Brook, NY 11794, USA}

%\author{Kiminad A. Mamo}\email{kamamo@wm.edu}
%\affiliation{Department of Physics, William \& Mary, Williamsburg, VA 23187, USA}

\author{Kiminad A. Mamo}
\thanks{\href{mailto:kamamo@wm.edu}{kamamo@wm.edu,}
\href{mailto:kamamo@jlab.org}{kamamo@jlab.org}}
\affiliation{Department of Physics, William and Mary, Williamsburg VA 23187, USA}
\affiliation{Theory Center, Jefferson Lab, Newport News, VA 23606, USA}

\author{Ismail Zahed}\email{ismail.zahed@stonybrook.edu}
\affiliation{Center for Nuclear Theory, Stony Brook University, Stony Brook, NY 11794, USA}

\date{\today}

\begin{abstract}
We revisit the two\texorpdfstring{‑}{-}dimensional Fourier transform of generalized parton distributions (GPDs) at nonzero skewness.
At $\eta=0$ it reduces to the standard impact\texorpdfstring{‑}{-}parameter density, while at $\eta\neq 0$ it is an \emph{off\texorpdfstring{‑}{-}forward} amplitude that we interpret as a genuine parton--nucleon correlation.
Its overall strength (the transverse-plane integral of the density) is fixed by the GPD at the kinematic point $t=-c_\eta=-4\eta^2 m_N^2/(1-\eta^2)$ and decreases monotonically with the rapidity gap
$\Delta y=\ln\!\bigl[(1+\eta)/(1-\eta)\bigr]=2\,\artanh(\eta)$.
This rapidity dependence implies \emph{rapidity\texorpdfstring{‑}{-}modified Ji identities} that connect helicity, orbital, and total angular momenta of the correlation in closed form.
To quantify these effects, we construct leading\texorpdfstring{‑}{-}twist quark and gluon GPDs in a string\texorpdfstring{‑}{-}based conformal framework:
conformal moments are parametrized by linear open\texorpdfstring{‑}{-} and closed\texorpdfstring{‑}{-}string Regge trajectories with slopes constrained by PDFs, hadron/glueball spectroscopy, and form\texorpdfstring{‑}{-}factor data, and GPDs are reconstructed over the full $(x,\eta,t)$ domain by Mellin--Barnes inversion with next-to-leading order (NLO) evolution.
We find qualitative agreement (and fair quantitative agreement within quoted uncertainties) for several moments and selected non\texorpdfstring{‑}{-}singlet $x$-space channels at $\mu=2\,$GeV when compared with lattice QCD, while we also identify channels with visible tension and discuss likely sources (PDF priors and $t$-slope systematics).
\end{abstract}

\maketitle
% \tableofcontents  % uncomment for preprint readability

% ======================================================================
\section{Introduction}
\label{sec:intro}

Generalized parton distributions (GPDs) provide a unified description of nucleon structure that interpolates between ordinary parton distribution functions (PDFs) and elastic form factors~\cite{Muller:1994ses,Ji:1996ek,Radyushkin:1996nd,Radyushkin:1997ki,Ji:1998xh,Diehl:2003ny,Belitsky:2005qn}.
They enter hard exclusive reactions such as deeply virtual Compton scattering (DVCS) and hard exclusive meson production, and will be constrained with increasing precision at Jefferson Lab and at the future Electron--Ion Collider (EIC)~\cite{Accardi:2012qut,Accardi:2023chb}.
A central appeal of GPDs is that they correlate longitudinal momentum fractions, transverse spatial distributions, and spin degrees of freedom within a Lorentz-covariant framework.

At zero skewness ($\eta=0$), the two-dimensional Fourier transform of the unpolarized GPD $H$ has a direct density interpretation in impact-parameter space~\cite{Soper:1976jc,Burkardt:2002hr}.
For $\eta\neq0$, the same transform does \emph{not} define a probability density; it is an off-diagonal matrix element of a gauge-invariant light-ray operator between distinct nucleon momentum eigenstates~\cite{Belitsky:2005qn,Diehl:2002he}.
In this paper, we sharpen this point and give it a concrete dynamical interpretation in the string-based conformal representation developed in Refs.~\cite{Mamo:2024jwp,Mamo:2024vjh,Hechenberger:2025wnz}.
In that approach, the $t$-channel is probed by color-singlet quark and gluon pairs described by linear open- and closed-string Regge trajectories~\cite{Brower:2006ea,Brower:2008cy}.
The recoiling nucleon carries a shifted longitudinal momentum, implying a rapidity gap
\begin{equation}
\Delta y \equiv |y_1-y_2| = \ln\!\Bigl(\frac{1+\eta}{1-\eta}\Bigr)=2\,\artanh(\eta),
\label{eq:rapidity_gap}
\end{equation}
whose derivation is recalled in Appendix~\ref{app:kin}.
At finite $\eta$, the transverse Fourier transform therefore probes an overlap between the nucleon and the exchanged color-singlet parton pair at transverse separation $\bperp$ \emph{and} rapidity gap $\Delta y$.

A key consequence is that the familiar Ji relations~\cite{Ji:1997gm} must be interpreted carefully at finite $\eta$.
Because the overall strength of the off-forward correlation depends on $\eta$ (equivalently on $\Delta y$), then the helicity, orbital, and total angular momenta extracted from twist-2 moments acquire explicit rapidity-dependent prefactors.
We show that these prefactors can be written in closed form and lead to rapidity-modified Ji identities that reduce to the standard ones at $\eta=0$.

From a practical perspective, conformal moments control Compton form factors in DVCS. Therefore, a faithful description of finite-skewness data requires a consistent treatment of the $\eta$ dependence already at the level of the conformal moments (see, e.g., Ref.~\cite{Kumericki:2016ehc} for the mapping between conformal moments and Compton form factors in global analyses).

On the phenomenological side, we reconstruct the leading-twist GPDs $H$, $E$, and $\widetilde H$ for quarks and gluons over the full kinematical domain in parton $x$, skewness $\eta$, and squared momentum transfer $t$, using the Mellin--Barnes resummation of conformal partial waves.
Our conformal moments are fixed at the hadronic scale $\mu_0=1\,$GeV by empirical PDFs and linear Regge slopes calibrated to hadron/glueball spectroscopy and elastic form factors, with no additional tunable shape parameters beyond these external inputs~\cite{Mamo:2024jwp,Mamo:2024vjh}.
After next-to-leading order (NLO) evolution governed by the Dokshitzer-Gribov-Lipatov-Altarelli-Parisi (DGLAP) and Efremov-Radyushkin-Brodsky-Lepage (ERBL) equations to $\mu=2\,$GeV, we compare moments and selected $x$-space channels with lattice QCD and discuss remaining tensions.

The paper is organized as follows.
Sec.~\ref{sec:setup} fixes conventions, kinematics, and the basic definitions needed later.
Sec.~\ref{sec:Moments} summarizes the Mellin--Barnes representation and our string-based parametrization of conformal moments, including the finite-skewness kernel and the $t$-slope inputs.
Sec.~\ref{sec:SpatialTomography} discusses spatial tomography at arbitrary skewness and the $\eta$-dependent norm of the parton--nucleon correlation.
Sec.~\ref{sec:SpinTomography} presents the spin decomposition at finite skewness, introduces the rapidity-modified Ji identities, and provides numerical illustrations.
We conclude in Sec.~\ref{sec:Conclusion}.
Appendix~\ref{app:kin} summarizes the kinematics and the rapidity-gap relation~\eqref{eq:rapidity_gap}, and Appendix~\ref{app:SkewNorm} derives the norm relation used throughout.

% ======================================================================
\section{Conventions and basic definitions}
\label{sec:setup}

We work in a symmetric frame with initial and final nucleon momenta $p_1^\mu$ and $p_2^\mu$,
average momentum $P^\mu=\tfrac12(p_1^\mu+p_2^\mu)$, and momentum transfer $\Delta^\mu=p_2^\mu-p_1^\mu$ with invariant momentum transfer $t=\Delta^2<0$.
The skewness is defined in the usual way through light-cone components,
$\eta\equiv -\Delta^+/(2P^+)$, so that (for vanishing transverse average momentum $\bm P_\perp=0$)
\begin{equation}
p_1^+= (1+\eta)P^+,\qquad p_2^+=(1-\eta)P^+.
\end{equation}
For $\eta\neq 0$ and purely transverse momentum transfer $\Dperp$, kinematics implies
\begin{equation}
t = -\frac{\Dperp^2 + 4\eta^2 m_N^2}{1-\eta^2}
   \equiv -\left(\frac{\Dperp^2}{1-\eta^2}+c_\eta\right),
\quad
c_\eta \equiv \frac{4\eta^2 m_N^2}{1-\eta^2},
\label{eq:t_eta}
\end{equation}
so that the minimal $|t|$ at fixed $\eta$ occurs at $\Dperp=0$ and equals $t=-c_\eta$. The nucleon mass in \eqref{eq:t_eta} is $m_N=0.938$ GeV.

Throughout, we use the standard charge-conjugation combinations
$F^{(\pm)}(x,\eta,t;\mu)=F(x,\eta,t;\mu)\mp F(-x,\eta,t;\mu)$ for quark GPDs.
The corresponding conformal moments are denoted by $\mathbb F^{(\pm)}(j,\eta,t;\mu)$, where $j$ is the conformal spin analytically continued to the complex plane.
We focus on the leading-twist GPDs $H$, $E$, and $\widetilde H$ and their conformal moments $\Hc$, $\Ec$, and $\Ht$.
(All results below are for $\widetilde E$-independent combinations and do not require modeling $\widetilde E$ explicitly.)

\begin{figure}[t]
  \centering
  \subfloat[]{%
    \includegraphics[width=0.82\linewidth]{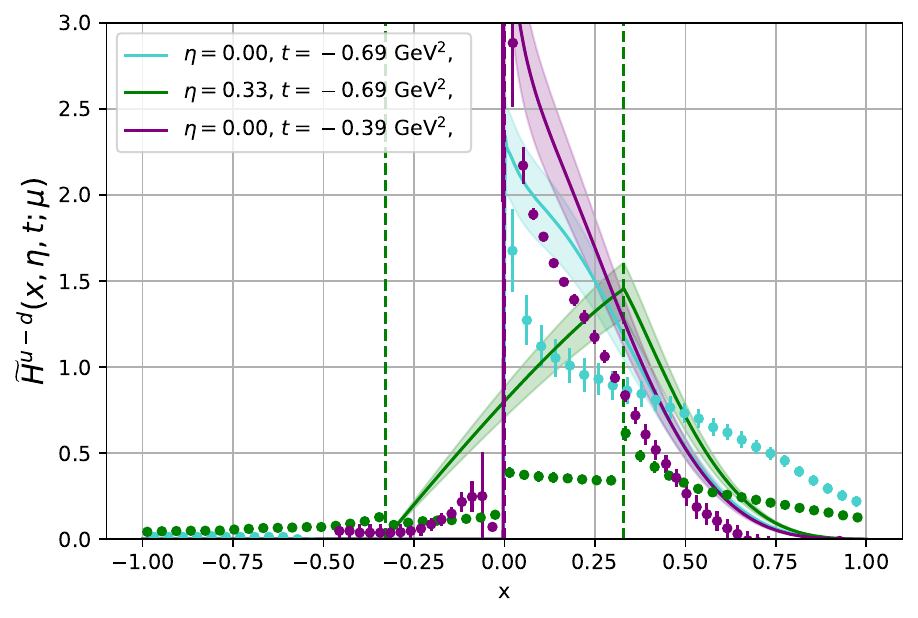}}\\[-2pt]
  \subfloat[]{%
    \includegraphics[width=0.82\linewidth]{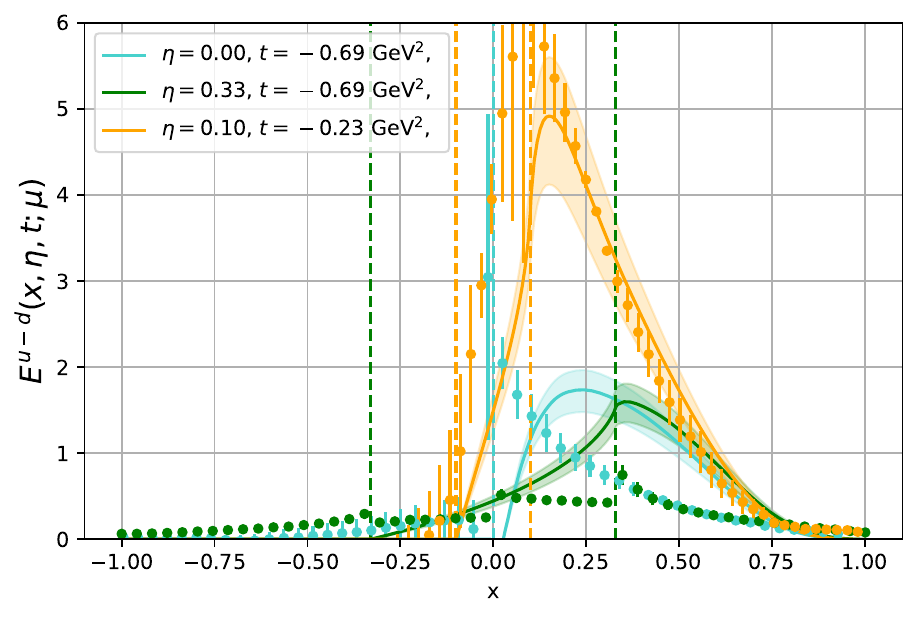}}
  \caption{\label{fig:GPDxspaceAll}
  Non-singlet/isovector helicity (a) and spin-flip (b) GPDs reconstructed from the MB representation~\eqref{eq:MB} using the conformal moments defined in Eqs.~\eqref{eq:ReggeMellin}--\eqref{eq:SkewnessRestore2}, compared with lattice data~\cite{Alexandrou:2020zbe,Holligan:2023jqh} (filled circles).
  Theory bands are at $\mu=2\,$GeV after NLO evolution.
  Vertical dashed lines mark $x=\pm\eta$ (ERBL--DGLAP boundaries).}
\end{figure}
% ======================================================================
\section{Conformal moments and string-based parametrization}
\label{sec:Moments}

\subsection{Mellin--Barnes representation}
\label{sec:MB}

Our starting point is the Mellin--Barnes (MB) representation of the leading-twist quark ($a=q$) and gluon ($a=g$) GPDs,
\begin{equation}
F_{a}(x,\eta,t;\mu)=\frac{s_a}{2\ii}
   \int_{c-\ii\infty}^{c+\ii\infty}\!
   \frac{\dd j}{\sin(\pi j)}\;
   p_{a}(j,x,\eta)\,
   \mathbb F_{a}(j,\eta,t;\mu),
\label{eq:MB}
\end{equation}
with $s_q=1$ and $s_g=-1$.
The kernels $p_{a}(j,x,\eta)$ are the analytically continued Gegenbauer partial waves in the ERBL ($|x|\le\eta$) and DGLAP ($x\ge\eta$) regions~\cite{Mueller:2005ed}.
The contour parameter $c$ is chosen so that all singularities of the integrand lie to the left of the contour, reproducing the conformal partial-wave expansion when the contour is closed.
All nontrivial hadronic dynamics are encoded in the conformal moments $\mathbb F_a(j,\eta,t;\mu)$, which are analytic functions of complex $j$ up to Regge-type singularities.

\subsection{Forward limit and Reggeized Mellin transform}
\label{sec:forward}

Gauge/string duality suggests that at high energy and fixed $t$ the dominant $t$-channel exchanges are approximately linear open- and closed-string trajectories~\cite{Brower:2006ea,Brower:2008cy}.
At the hadronic input scale $\mu_0=1\,$GeV and at $\eta=0$ we model the conformal moments by a ``Reggeized'' Mellin transform,
\begin{equation}
\mathcal F_{a}(j,t;\mu_0)
  \equiv \mathbb F_{a}(j,0,t;\mu_{0})
  =\int_{0}^{1}\!\dd x\;
      \frac{f_{a}(x,\mu_{0})}{x^{\,j-1+\alpha'_{a}t}},
\label{eq:ReggeMellin}
\end{equation}
where $f_a$ are the empirical forward PDFs at $\mu_0$ and $\alpha'_a$ are linear Regge slopes.
We use the MSTW09 NLO set for unpolarized PDFs~\cite{Martin:2009iq} and the AAC NLO set for polarized PDFs~\cite{Hirai:2006sr}, which allow analytic Mellin transforms in Eq.~\eqref{eq:ReggeMellin} and match the Regge calibrations used in Refs.~\cite{Mamo:2024jwp,Mamo:2024vjh}.

The slopes used in this work are collected in Table~\ref{tab:slopes}.
For some channels a secondary trajectory with slope $\alpha_a^{\prime\mathrm S}$ is introduced to cancel spurious poles in the finite-skewness continuation (see Sec.~\ref{sec:skewness}).

\begin{table*}[t]
\centering
\caption{\label{tab:slopes}
Open- and closed-string slopes (GeV$^{-2}$) entering the Reggeized Mellin transform~\eqref{eq:ReggeMellin}.
A dash denotes that no secondary trajectory is required in that channel.}
\begin{tabular}{lccccc}
\hline\hline
Channel & $\alpha'_{u-d}$ & $\alpha'_{u+d}$ & $\alpha'^{\mathrm S}_{u+d}$
        & $\alpha'_{g}$   & $\alpha^{\prime\mathrm S}_{g}$ \\ \hline
$\Hc^{(\pm)}$             & 0.66 & 0.96 & 1.83 & 0.63 & 4.28 \\
$\Ec^{(\pm)}$             & 1.48 & 1.12 &   -- &   -- &   -- \\
$\widetilde{\Hc}^{(\pm)}$ & 0.45 & 0.30 & 1.18 & 0.49 & 0.74 \\
\hline\hline
\end{tabular}
\end{table*}

\subsection{Finite skewness and analytic continuation}
\label{sec:skewness}

To extend the moments to $\eta\neq 0$ we use the unique hypergeometric kernel derived in cubic string field theory that enforces polynomiality, crossing symmetry, and removes spurious poles~\cite{Nishio:2014rya,Mamo:2024jwp}.\footnote{While finalizing this manuscript it was brought to our attention that aspects of the $\eta=0$ construction overlap with other conformal parametrizations (e.g.\ Ref.~\cite{Kumericki:2007sa}). In our framework the dynamical motivation and, crucially, the finite-skewness dependence through Eqs.~\eqref{eq:SkewnessRestore}--\eqref{eq:SkewnessRestore2} provide a clear discriminator.}
We then obtain:
\begin{equation}
\mathbb F_{a}(j,\eta,t;\mu_{0})=
 \bigl[\hat d_{j}(\eta,t)-1\bigr]
 \bigl[\mathcal F_{a}(j,t;\mu_0)-\mathcal F^{\mathrm S}_{a}(j,t;\mu_0)\bigr]
 +\mathcal F_{a}(j,t;\mu_0),
\label{eq:SkewnessRestore}
\end{equation}
with
\begin{equation}
\hat d_{j}(\eta,t)={}_{2}F_{1}\!\Bigl(\tfrac12-j,\,-\tfrac{j}{2};\tfrac12;
                                 -\tfrac{4m_{N}^{2}\eta^{2}}{t}\Bigr).
\label{eq:SkewnessRestore2}
\end{equation}
Here ${}_{2}F_{1}$ is the Gauss hypergeometric function and $\mathcal F^{\mathrm S}_{a}(j,t)$ denotes the secondary trajectory contribution obtained from Eq.~\eqref{eq:ReggeMellin} by $\alpha'_a\to \alpha_a^{\prime\mathrm S}$.
This construction guarantees the correct ERBL/DGLAP support and polynomiality of Mellin moments at any skewness, while retaining an analytic structure controlled by linear trajectories.

\subsection{Evolution and MB inversion}
\label{sec:evolMB}

All leading-twist singlet and non-singlet moments $\{\Hc,\Ec,\Ht\}$ are evolved from $\mu_0$ to a generic scale $\mu$ by solving the NLO DGLAP--ERBL evolution equations directly in conformal space.
The $x$-space GPDs at any $(x,\eta,t)$ are then obtained by numerical MB inversion of Eq.~\eqref{eq:MB}.

As an illustration, Fig.~\ref{fig:GPDxspaceAll} shows selected non-singlet/isovector channels reconstructed with a single Regge parameter set, compared to lattice results.
The orange curve is consistent within uncertainties, while the green curves (both panels) and the blue curve (bottom) exhibit visible tension.
Within our framework these deviations are driven primarily by the forward PDF priors and by residual systematics associated with the $t$-slopes entering Eq.~\eqref{eq:ReggeMellin}; we return to this in Sec.~\ref{sec:SpinTomography}.

\begin{figure}[t]
  \centering
  \subfloat[$\eta = 0$]{%
    \includegraphics[width=0.78\linewidth]{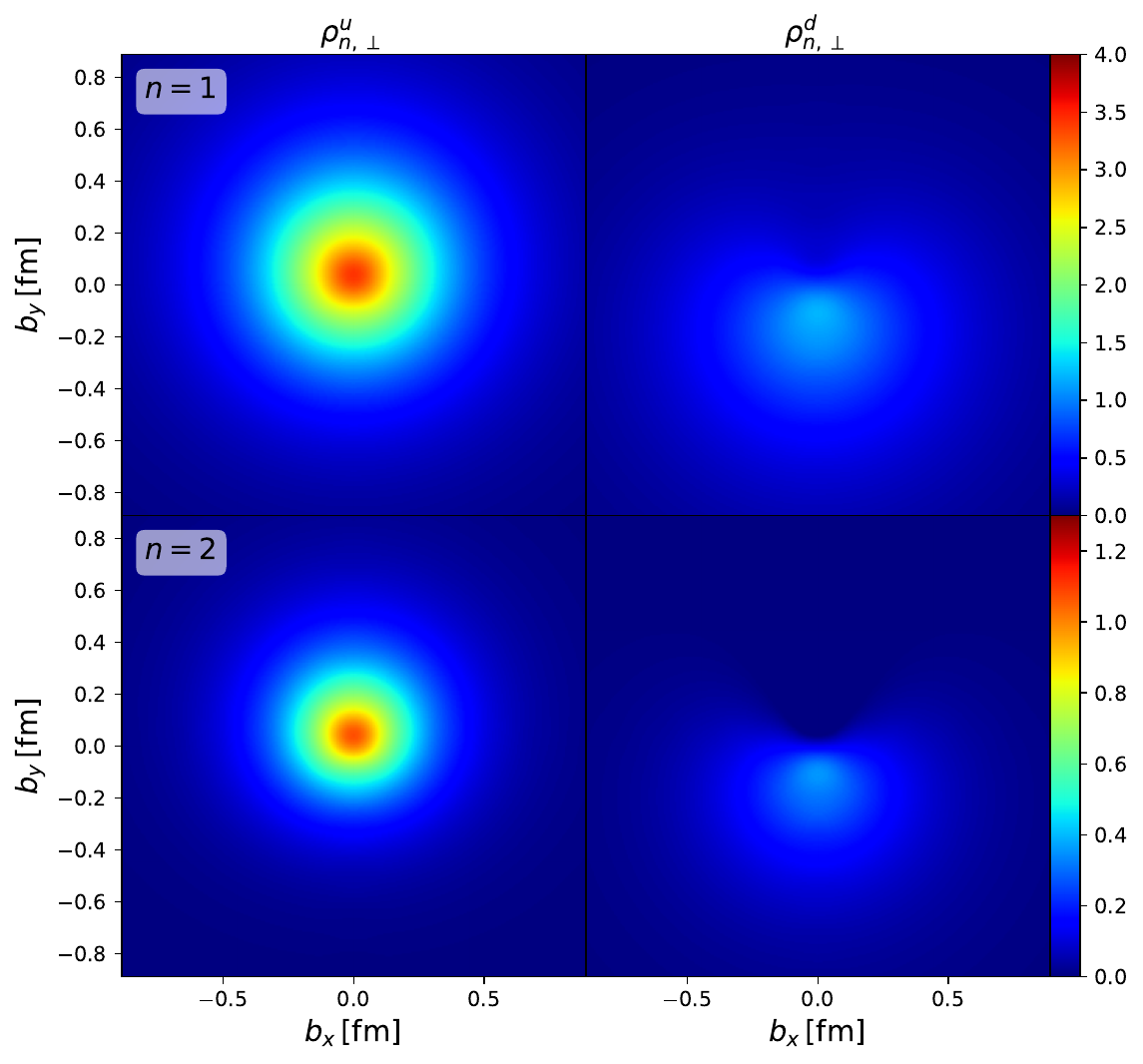}}\\[-2pt]
  \subfloat[$\eta = 0.33$]{%
    \includegraphics[width=0.78\linewidth]{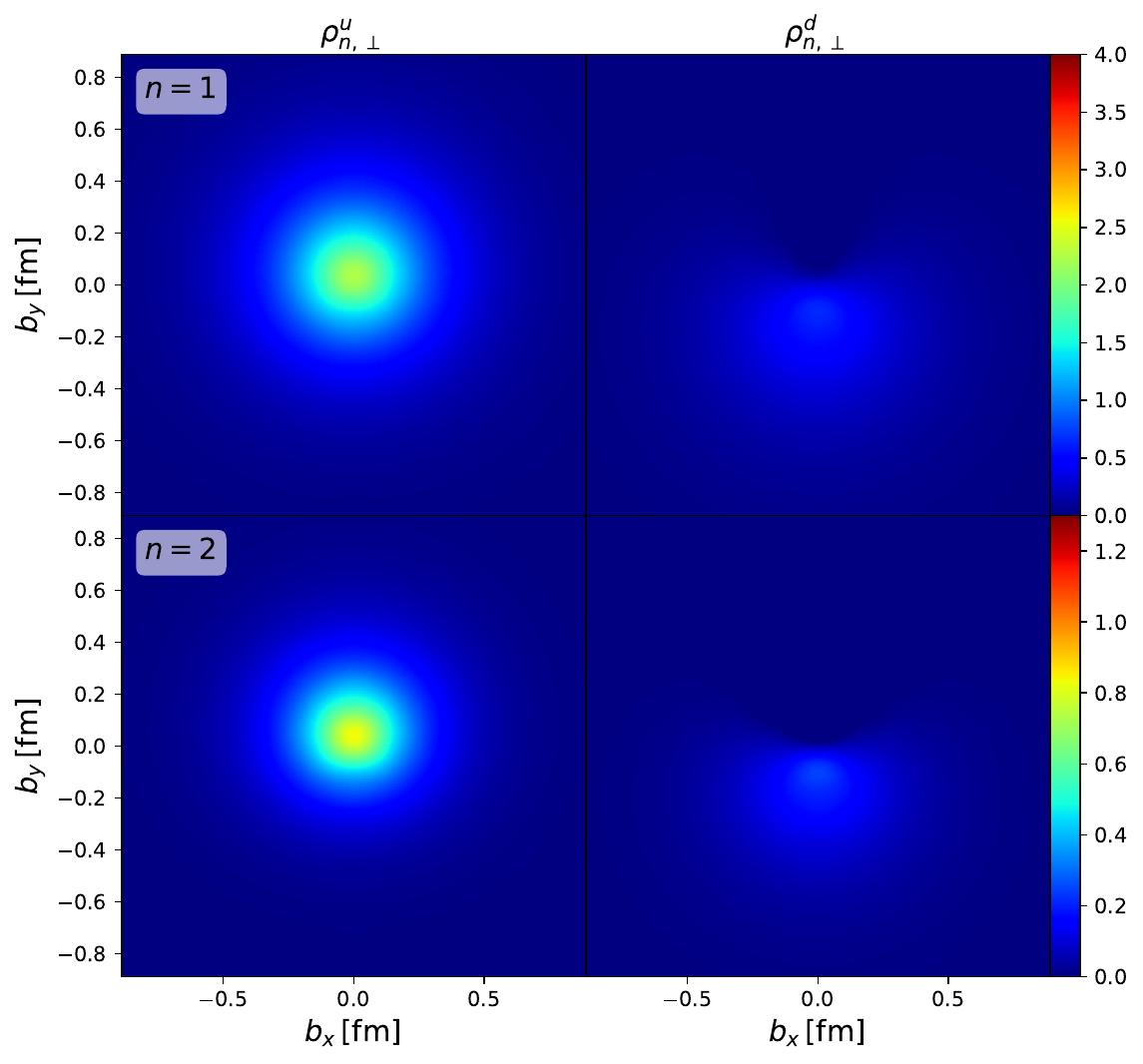}}
  \caption{\label{fig:ImpactMomentsCompare}
  Spatial tomography at $\mu=2$\,GeV for non-singlet transverse densities $\rho^{u/d}_{n}(\bperp;\eta)$ (with $n=1,2$) defined in Eq.~\eqref{eq:bspaceFT} for a nucleon polarized along $+\hat{\bm y}$.
  Panel (a) shows $\eta=0$ (impact-parameter densities).
  Panel (b) shows $\eta=0.33$, corresponding to a rapidity gap $\Delta y=2\,\artanh(\eta)\simeq 0.69$; the overall norms decrease according to Eq.~\eqref{eq:NormVariationFinal}.}
\end{figure}

% ======================================================================
\section{Spatial tomography at arbitrary skewness}
\label{sec:SpatialTomography}

\subsection{Impact-parameter densities and their interpretation}

For a nucleon polarized along $+\hat{\bm y}$, the two-dimensional Fourier transform of the conformal moments defines the transverse profile of the corresponding off-forward matrix element.
Here and below, superscripts on transverse vectors denote Cartesian components and should not be confused with the longitudinal rapidity $y$ used in Eq.~\eqref{eq:rapidity_gap}.
For non-singlet flavor combinations one convenient choice is
\begin{align}
\rho^{u\pm d}_{n}(\bperp;\eta;\mu)
   &=\!\int\!\frac{\dd^{2}\Dperp}{(2\pi)^{2}}\,
       e^{- \ii\,\Dperp\cdot\bperp}\,
       \Bigl[
         \mathbb{H}^{(-)}_{u\pm d}\!\bigl(n,\eta,t;\mu\bigr)
         \nonumber\\
         &+\,\ii\,\frac{\Delta_{\perp}^{\,y}}{2m_{N}}\,
            \mathbb{E}^{(-)}_{u\pm d}\!\bigl(n,\eta,t;\mu\bigr)
       \Bigr],
\label{eq:bspaceFT}
\end{align}
Flavor-separated densities are obtained in the usual way from isoscalar/isovector combinations,
\begin{equation}
\rho^{u}_{n}=\tfrac12\bigl(\rho^{u+d}_{n}+\rho^{u-d}_{n}\bigr),
\qquad
\rho^{d}_{n}=\tfrac12\bigl(\rho^{u+d}_{n}-\rho^{u-d}_{n}\bigr).
\end{equation}
where $n=1,2,\ldots$ labels the conformal spin (integer moments), $\Delta_\perp^{\,y}$ is the $y$-component of $\Dperp$, and $t$ is related to $(\Dperp,\eta)$ by Eq.~\eqref{eq:t_eta}.
At $\eta=0$, $\rho_n$ reduces to the familiar impact-parameter density associated with the $n$th conformal moment~\cite{Soper:1976jc,Burkardt:2002hr}.
For $\eta\neq 0$, $\rho_n$ should instead be viewed as a genuine \emph{parton--nucleon correlation} amplitude in transverse space, since it connects nucleon states with different longitudinal momenta.

\subsection{Norm of the correlation and rapidity dependence}

The overall strength (norm) of the correlation is obtained by integrating Eq.~\eqref{eq:bspaceFT} over the transverse plane,
\begin{equation}
   N^{u\pm d}_{n}(\eta;\mu)\equiv\!\int\!\dd^{2}\bperp\,\rho^{u\pm d}_{n}(\bperp;\eta;\mu)
   =\mathbb H_{u\pm d}^{(-)}\bigl(n,\eta,t=-c_\eta;\mu\bigr),
\label{eq:NormVariationFinal}
\end{equation}
with $c_\eta$ defined in Eq.~\eqref{eq:t_eta}.%
\footnote{For notational simplicity we display the relation for the non-singlet combinations used in Eq.~\eqref{eq:bspaceFT}. The derivation in Appendix~\ref{app:SkewNorm} is completely general and applies to singlet and gluon channels as well. In particular, for the $C$-even singlet combination we write $N_{n}^{(+)}(\eta)\equiv \mathbb H^{(+)}(n,\eta,-c_\eta)$; this is the quantity shown in Fig.~\ref{fig:Rapidity4panel}(d) for $n=2$.}
Eq.~\eqref{eq:NormVariationFinal} follows directly from the Hankel representation of the Fourier transform; we provide a compact derivation in Appendix~\ref{app:SkewNorm}.
Two features are worth emphasizing:
(i) at fixed $\eta$ the relevant ``forward'' point in the $t$-channel is not $t=0$ but the kinematic point $t=-c_\eta$ corresponding to $\Dperp=0$; and
(ii) as $|\eta|$ (equivalently $\Delta y$) increases, $c_\eta$ increases and the norm decreases because the underlying color-singlet probe becomes less correlated with the nucleon at large rapidity gaps.

Fig.~\ref{fig:ImpactMomentsCompare} illustrates this behavior for representative non-singlet conformal moments in transverse space.

\begin{figure}[t]
  \centering
  \subfloat[\textbf{Gluon helicity}]{\includegraphics[width=.46\columnwidth]{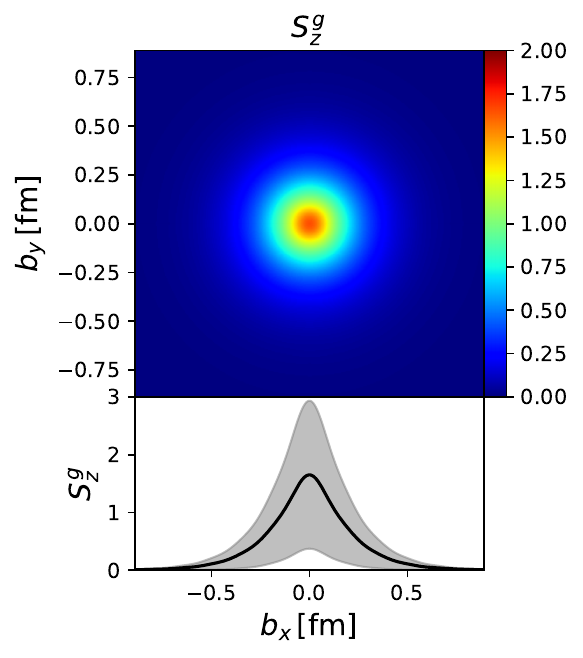}}\hspace{0.04\columnwidth}
  \subfloat[\textbf{Sea-quark helicity}]{\includegraphics[width=.46\columnwidth]{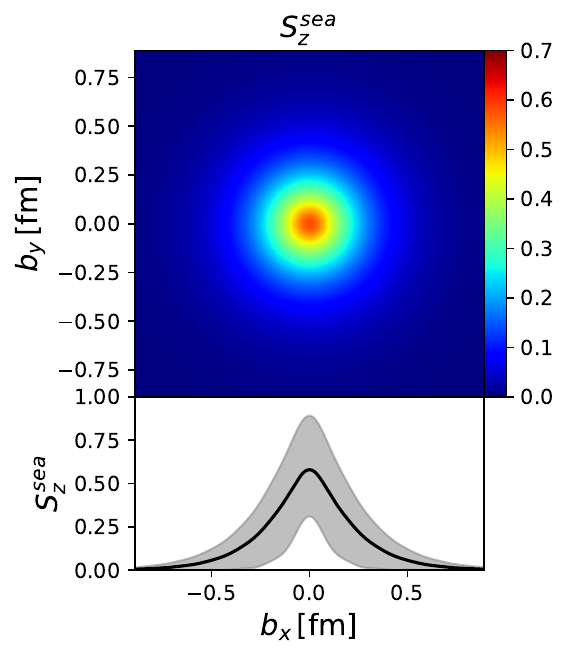}}\\[4pt]
  \subfloat[\textbf{Up-quark helicity}]{\includegraphics[width=.46\columnwidth]{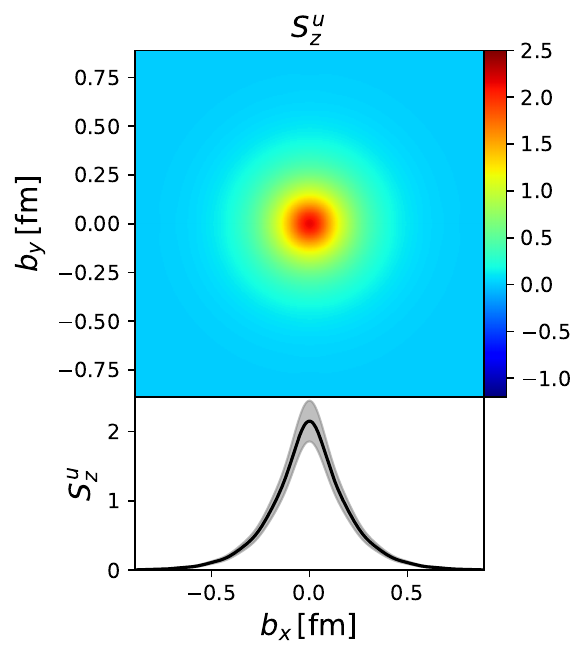}}\hspace{0.04\columnwidth}
  \subfloat[\textbf{Down-quark helicity}]{\includegraphics[width=.46\columnwidth]{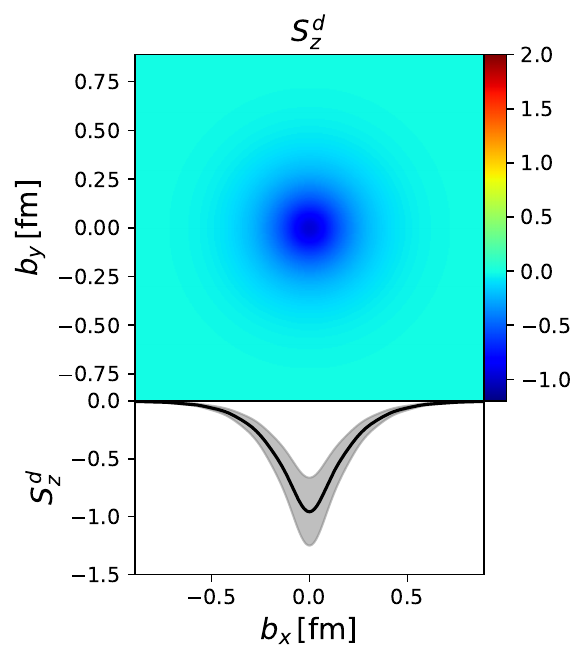}}
  \caption{\label{fig:SpinDecomp_helicity}
  Impact-parameter distributions for gluon (a), sea- (b), up- (c), and down- (d) quark helicity densities at $\mu=2$\,GeV and $\eta=0$ obtained from Eq.~\eqref{eq:Ji_decomp}.}
\end{figure}

\begin{figure}[t]
  \centering
  \subfloat[\textbf{Up-quark OAM}]{\includegraphics[width=.46\columnwidth]{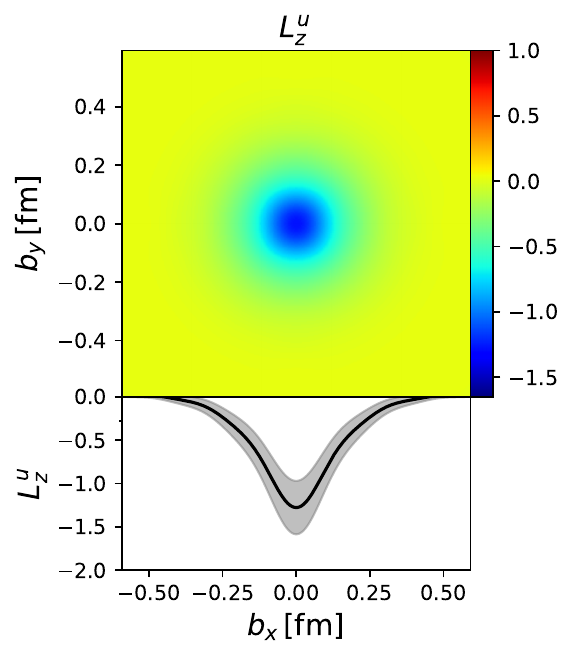}}\hspace{0.04\columnwidth}
  \subfloat[\textbf{Down-quark OAM}]{\includegraphics[width=.46\columnwidth]{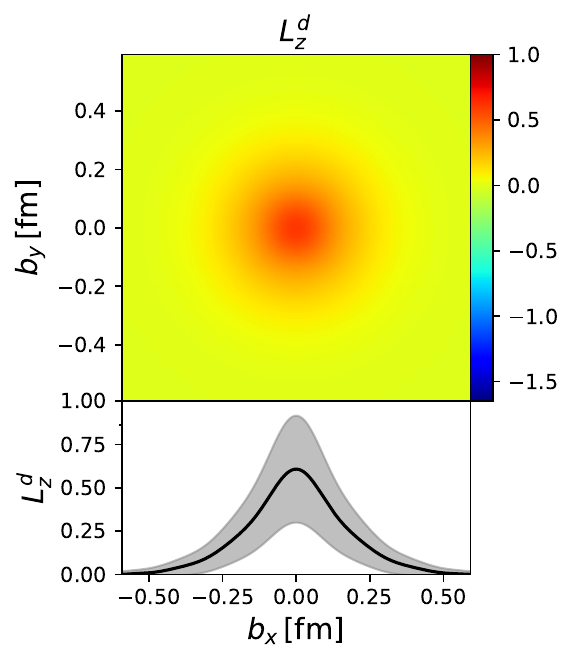}}\\[4pt]
  \subfloat[\textbf{Up-quark $C_{z}^u$}]{\includegraphics[width=.46\columnwidth]{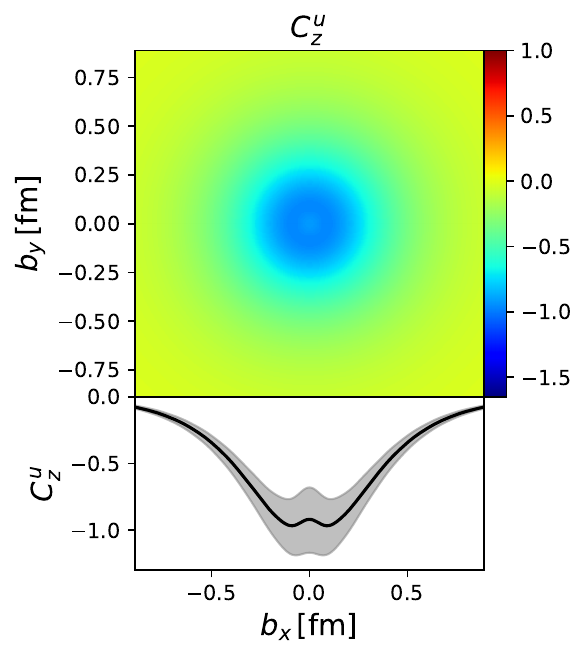}}\hspace{0.04\columnwidth}
  \subfloat[\textbf{Down-quark $C_{z}^d$}]{\includegraphics[width=.46\columnwidth]{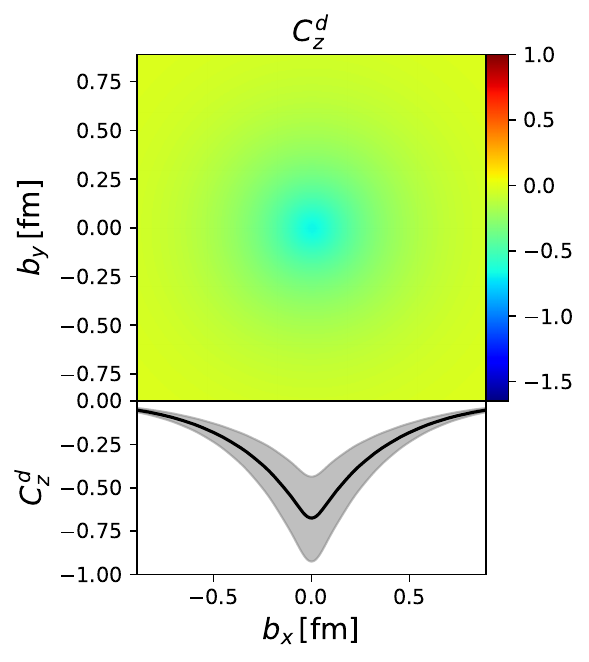}}
  \caption{\label{fig:SpinDecomp_oam}
  Impact-parameter distributions for up- (a) and down-quark (b) orbital angular momentum and up- (c) and down-quark (d) spin--orbit correlations at $\mu=2$\,GeV and $\eta=0$.
  The densities follow from Eqs.~\eqref{eq:Ji_decomp} and~\eqref{eq:spin_orbit}.}
\end{figure}

\begin{figure}[t]
  \centering
  \subfloat[Spin--orbit correlation $C_{z}^{u+d}$]{%
    \includegraphics[width=0.82\linewidth]{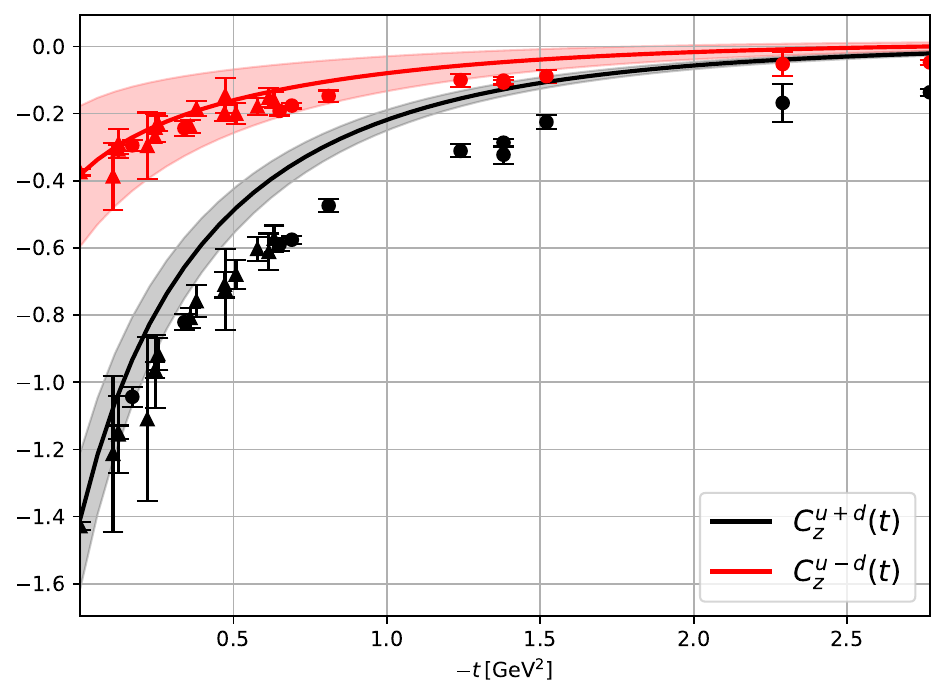}}\\[-2pt]
  \subfloat[Orbital angular momentum $L_{z}^{u+d}$]{%
    \includegraphics[width=0.82\linewidth]{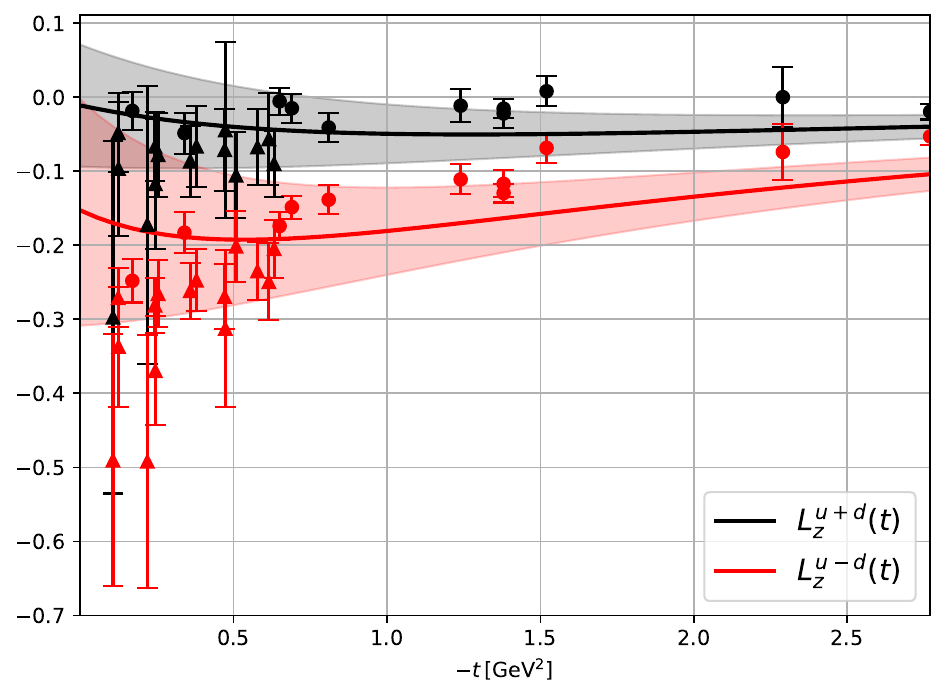}}
  \caption{\label{fig:SpinOrbital_t}
  Isoscalar spin--orbit correlation (a) and quark orbital angular momentum (b) versus $-t$ at $\mu=2$\,GeV, computed from Eqs.~\eqref{eq:Ji_decomp} and~\eqref{eq:spin_orbit} at $\eta=0$.
  Bands: this work (PDF-propagated uncertainties).
  Points: lattice results~\cite{Bhattacharya:2024wtg,LHPC:2007blg}.}
\end{figure}

% ======================================================================
\section{Spin tomography and rapidity-modified Ji identities}
\label{sec:SpinTomography}

\subsection{Ji decomposition from conformal moments}

For any $(\eta,t)$, the axial moment $\Ht_a$ governs the helicity carried by quarks ($a=q$) or gluons ($a=g$), while the vector moments $\Hc_a$ and $\Ec_a$ govern the total angular momentum.
In our conventions, the Ji decomposition reads
\begin{subequations}\label{eq:Ji_decomp}
\begin{align}
  S^{z}_{a}(\eta,t;\mu)
    &=\tfrac{1}{2}\,\Ht_{a}^{(+)}(1,\eta,t;\mu),
\\
  \mathcal{J}^{z}_{a}(\eta,t;\mu)
    &=\tfrac{1}{2}\Bigl[\Hc_{a}^{(+)}(2,\eta,t;\mu)+\Ec_{a}^{(+)}(2,\eta,t;\mu)\Bigr],
\\
  L^{z}_{a}(\eta,t;\mu)
    &=\mathcal{J}^{z}_{a}(\eta,t;\mu)-S^{z}_{a}(\eta,t;\mu).
\end{align}
\end{subequations}
At $\eta=0$ and $t=0$ these reduce to the familiar spin sum rule $\sum_{a=q,g}\mathcal{J}^z_a=\tfrac12$~\cite{Ji:1997gm}.
We also consider the spin--orbit correlation~\cite{Lorce:2014mxa},
\begin{equation}
  C^{z}_{a}(\eta,t;\mu)=\tfrac{1}{2}\Bigl[\,\Ht_{a}^{(-)}(2,\eta,t;\mu)
                                    -\Hc_{a}^{(-)}(1,\eta,t;\mu)\Bigr],
\label{eq:spin_orbit}
\end{equation}
where we neglect the contribution proportional to the quark mass $m_q$ from the transversity GPDs $\widetilde{H}^a_T$ and $E_T^a$, respectively.
\begin{figure*}[t]
  \centering
  \subfloat[\textbf{Total spin} $\mathcal{J}_{z}(\eta)$]{%
      \includegraphics[width=.41\linewidth]{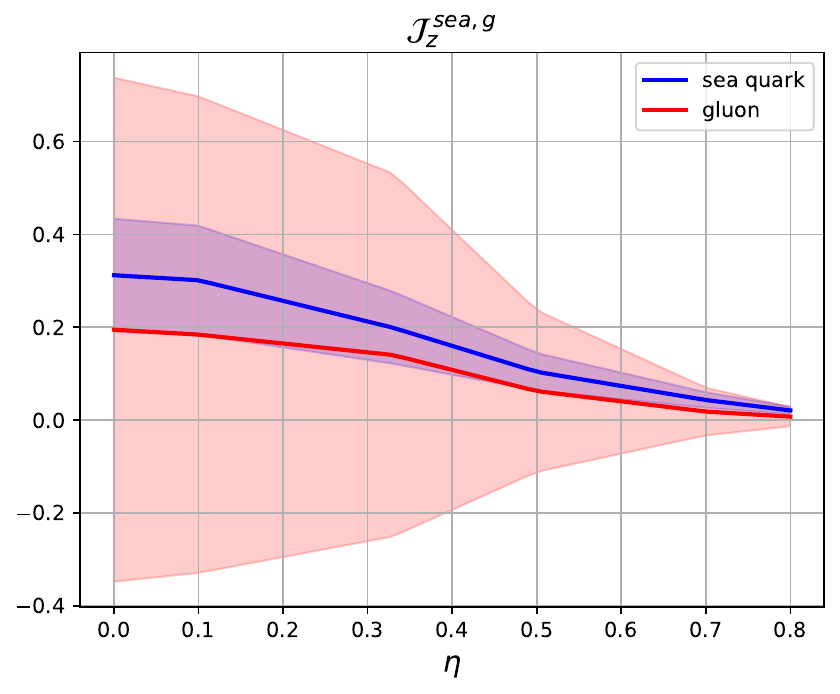}}
  \hspace{0.04\linewidth}
  \subfloat[\textbf{Helicity} $S_{z}(\eta)$]{%
      \includegraphics[width=.41\linewidth]{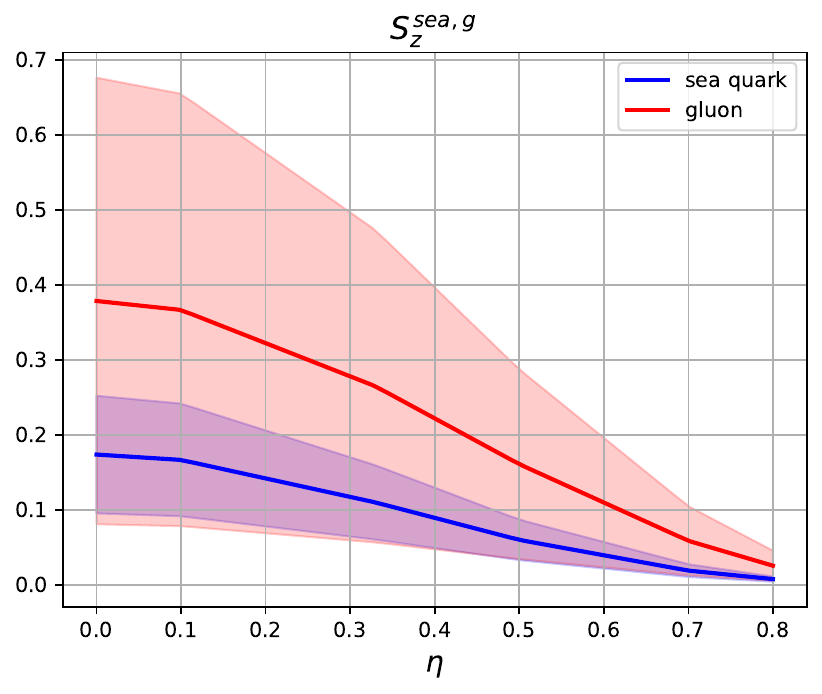}}\\[6pt]
  \subfloat[\textbf{OAM} $ L_{z}(\eta)$]{%
      \includegraphics[width=.41\linewidth]{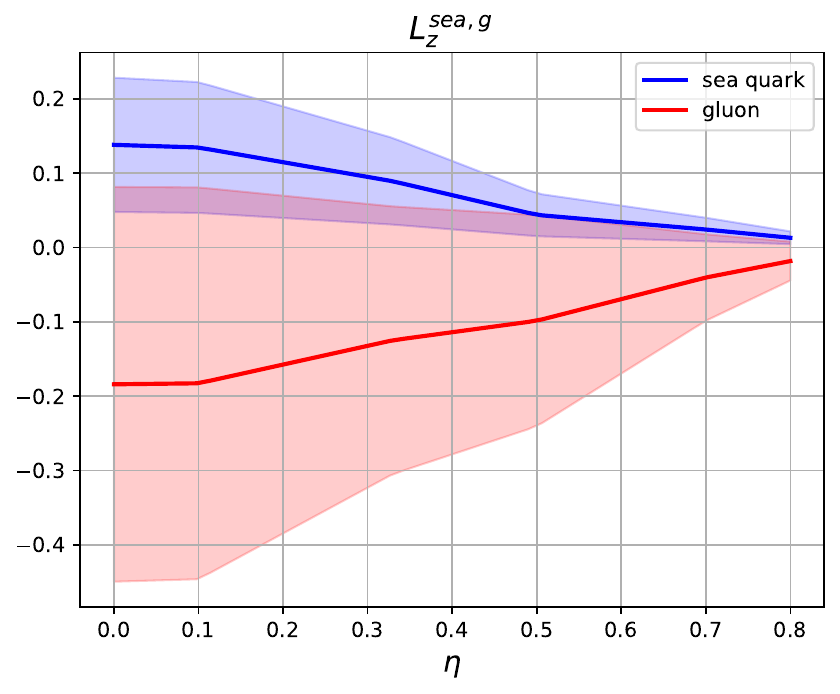}}
  \hspace{0.04\linewidth}
  \subfloat[\textbf{Norm} $N_{2}^{(+)}(\eta)$]{%
      \includegraphics[width=.41\linewidth]{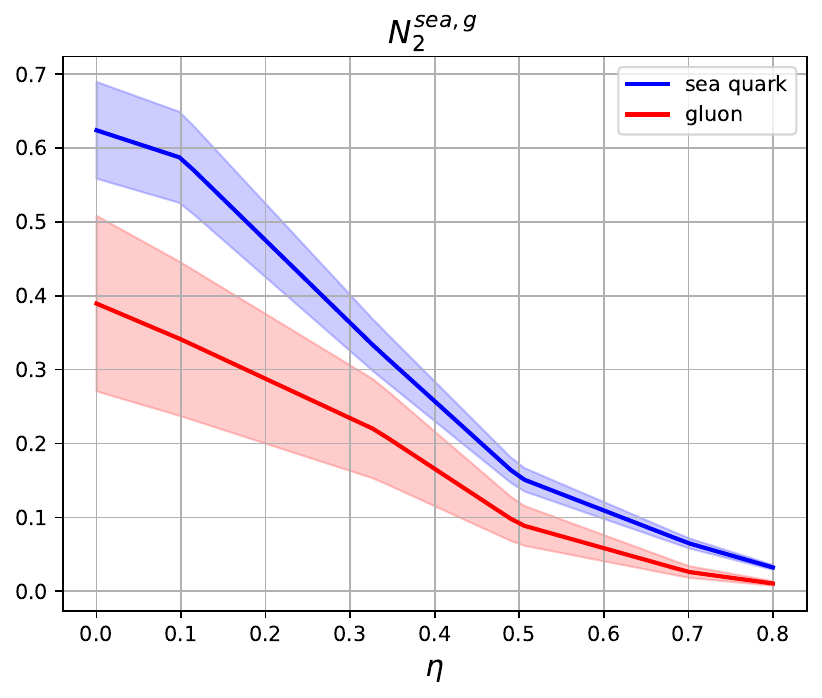}}
  \caption{\label{fig:Rapidity4panel}
  Rapidity dependence at $\mu=2$\,GeV of (a) total spin $\mathcal{J}_z(\eta)$, (b) helicity $S_z(\eta)$, (c) orbital angular momentum $L_z(\eta)$, and (d) the correlation norm $N_{2}^{(+)}(\eta)=\mathbb H^{(+)}(2,\eta,-c_\eta)$ (cf.\ Eq.~\eqref{eq:NormVariationFinal}).
  Spin observables are defined at $t=-c_\eta$ through Eq.~\eqref{eq:effective_spin}.
  The horizontal axis is the rapidity gap $\Delta y$ related to $\eta$ by Eq.~\eqref{eq:rapidity_gap}.
  Theory bands denote propagated PDF uncertainties.}
\end{figure*}

\subsection{Spin budget and transverse densities at $\eta=0$}

Table~\ref{tab:SpinBudgetCombined} summarizes our proton spin decomposition at $\eta=0$ and $t=0$ for $\mu_0=1\,$GeV (input) and after NLO evolution to $\mu=2\,$GeV, along with representative lattice determinations of quark helicities.
Fig.~\ref{fig:SpinDecomp_helicity} shows the corresponding impact-parameter distributions of helicity densities at $\eta=0$, and Fig.~\ref{fig:SpinDecomp_oam} shows the orbital-angular-momentum and spin--orbit densities.
Finally, Fig.~\ref{fig:SpinOrbital_t} displays the $t$-dependence of $L_z$ and $C_z$ and compares it to available lattice data.

\begin{table*}[t]
\centering
\begin{tabular}{lccccc}
\hline\hline
\multicolumn{6}{c}{\textbf{(a) $\mu = 2\,$GeV (NLO evolved)}}\\
\hline
                       & $u$            & $d$             & $s$            & $g$             & total \\
$S_{z}$ (this work)              &  0.421(66)    & $-0.210(66)$  & $-0.039(194)$ & 0.372(330)     & 0.544(394) \\
$S_{z}$ \cite{Alexandrou:2020sml} &  0.432(8)     & $-0.213(8)$   & $-0.023(4)$   & --             & --          \\
$S_{z}$ \cite{Liang:2018pis}      &  0.424(18)    & $-0.204(12)$  & $-0.018(4)$   & --             & --          \\
$S_{z}$ \cite{Green:2017keo}      &  0.432(7)     & $-0.173(54)$  & $-0.012(12)$  & --             & --          \\[2pt]
$L_{z}$ (this work)              & $-0.082(86)$  &  0.071(86)    &  0.147(217)   & $-0.179(340)$  & $-0.044(421)$ \\[2pt]
$\mathcal{J}_{z}$ (this work)    &  0.339(110)   & $-0.139(110)$ &  0.108(291)   & 0.192(474)     & 0.50(57) \\
\hline
\multicolumn{6}{c}{\textbf{(b) $\mu = 1\,$GeV (input scale)}}\\
\hline
                       & $u$            & $d$             & $s$            & $g$             & total \\
$S_{z}$ (this work)              &  0.421(66)    & $-0.210(66)$  & $-0.027(185)$ & 0.156(185)     & 0.341(280) \\
$L_{z}$ (this work)              & $-0.071(86)$  &  0.066(86)    &  0.141(210)   & 0.023(206)     & 0.159(318) \\
$\mathcal{J}_{z}$ (this work)    &  0.350(110)   & $-0.143(110)$ &  0.114(280)   & 0.179(279)     & 0.50(42) \\
\hline\hline
\end{tabular}
\caption{\label{tab:SpinBudgetCombined}
Proton spin decomposition at $\eta=0$ and $t=0$ based on Eqs.~\eqref{eq:Ji_decomp} and~\eqref{eq:spin_orbit}.
Uncertainties in parentheses are propagated from the input PDFs.
For comparison we list representative lattice results for quark helicities (gluons not quoted in those works).}
\end{table*}

\subsection{Finite skewness: rapidity-modified Ji identities}

At $\eta\neq 0$, the transverse Fourier transform defines an off-forward correlation whose norm varies with $\eta$ according to Eq.~\eqref{eq:NormVariationFinal}.
To connect spin observables to this correlation, we evaluate the moments at $\Dperp=0$, i.e.\ at the kinematic point $t=-c_\eta$.
We define the corresponding rapidity-dependent (effective) contributions
\begin{subequations}\label{eq:effective_spin}
\begin{align}
  S_{z}^{a}(\eta;\mu)&\equiv S^{z}_{a}(\eta,-c_\eta;\mu)=\tfrac12\,\Ht_{a}^{(+)}(1,\eta,-c_\eta;\mu),
\\
  \mathcal{J}_{z}^{a}(\eta;\mu)&\equiv \mathcal{J}^{z}_{a}(\eta,-c_\eta;\mu)
   =\tfrac12\bigl[\Hc_{a}^{(+)}(2,\eta,-c_\eta;\mu)\nonumber\\
   &\qquad\qquad\qquad\qquad+\Ec_{a}^{(+)}(2,\eta,-c_\eta;\mu)\bigr],
\\
  L_{z}^{a}(\eta;\mu)&\equiv L^{z}_{a}(\eta,-c_\eta;\mu)=\mathcal{J}_{z}^{a}(\eta;\mu)-S_{z}^{a}(\eta;\mu).
\end{align}
\end{subequations}

For the second moment, Lorentz covariance and polynomiality relate the conformal moments to the nucleon matrix elements of the (quark/gluon) energy--momentum tensor.
In a standard parametrization one introduces gravitational form factors $A_a(t)$ and $B_a(t)$ and a $D$-term form factor $D_a(t)$ such that~\cite{Belitsky:2005qn,Ji:1997gm}
\begin{subequations}\label{eq:poly_j2}
\begin{align}
\Hc_{a}^{(+)}(2,\eta,t;\mu) &= A_a(t;\mu) + \eta^2 D_a(t;\mu),
\\
\Ec_{a}^{(+)}(2,\eta,t;\mu) &= B_a(t;\mu) - \eta^2 D_a(t;\mu),
\end{align}
\end{subequations}
so that the $D$-term cancels in the sum and
\begin{equation}
\Hc_{a}^{(+)}(2,\eta,t)+\Ec_{a}^{(+)}(2,\eta,t)=A_a(t)+B_a(t),
\label{eq:AB_from_HE}
\end{equation}
independent of $\eta$.

Evaluating the correlation at $\Dperp=0$ therefore amounts to evaluating the form factors at $t=-c_\eta$, which decreases the total spin with increasing rapidity gap.
Summing over quark and gluon channels gives the rapidity-modified Ji identity
\begin{align}
\mathcal{J}_z(\eta;\mu)&\equiv\sum_{a=q,g}\mathcal{J}_z^a(\eta;\mu)
=\sum_{a=q,g}\bigl[S_z^a(\eta;\mu)+L_z^a(\eta;\mu)\bigr]
\nonumber\\
&=\tfrac12\sum_{a=q,g}\Bigl[\Hc_{a}^{(+)}(2,\eta,-c_\eta;\mu)+\Ec_{a}^{(+)}(2,\eta,-c_\eta;\mu)\Bigr]\nonumber\\
&=\tfrac12\bigl[A(-c_\eta)+B(-c_\eta)\bigr],
\label{eq:rapidity_Ji}
\end{align}
with $A(t)\equiv A_q(t;\mu)+A_g(t;\mu)$ and $B(t)\equiv B_q(t;\mu)+B_g(t;\mu)$, both independent of the scale $\mu$.
At $t=0$ one has $A(0)=1$ and $B(0)=0$ (vanishing anomalous gravitomagnetic moment)~\cite{Teryaev:1999su,Brodsky:2000ii}, so that $\mathcal{J}_z(0)=\tfrac12$.

Fig.~\ref{fig:Rapidity4panel} shows the resulting rapidity dependence at $\mu=2\,$GeV.
The depletion at large $\Delta y$ is a direct reflection of the reduced parton--nucleon overlap encoded in Eq.~\eqref{eq:NormVariationFinal} and is qualitatively consistent with the suppression of string-mediated exchanges at large rapidity gaps in the Regge limit~\cite{Basar:2012jb,Liu:2018gae}.

% ======================================================================
\section{Conclusions}
\label{sec:Conclusion}

We presented a rapidity-resolved interpretation of transverse Fourier transforms of GPDs at nonzero skewness.
While the $\eta=0$ transform yields an impact-parameter density, at $\eta\neq0$ it defines an off-forward parton--nucleon correlation whose overall strength is fixed by the GPD at the kinematic point $t=-c_\eta$ and decreases with the rapidity gap $\Delta y=2\,\artanh(\eta)$.
This norm variation induces rapidity-dependent prefactors in the spin decomposition and leads to rapidity-modified Ji identities, Eq.~\eqref{eq:rapidity_Ji}, that reduce to the standard Ji relations at $\eta=0$.

Quantitatively, we reconstructed leading-twist quark and gluon GPDs $H$, $E$, and $\widetilde H$ over the full $(x,\eta,t)$ domain using a string-based conformal parametrization:
linear open- and closed-string trajectories determine the $t$-dependence, empirical PDFs determine the forward limits, and a hypergeometric kernel enforces polynomiality and crossing symmetry at finite skewness.
After NLO evolution to $\mu=2\,$GeV, we obtained qualitative agreement with lattice QCD for several moments and selected non-singlet $x$-space channels, while we also identified channels with visible tension that most plausibly originate from PDF priors and residual slope systematics.
Because the framework is analytic, closed under evolution, and fast to evaluate, it is well suited for global DVCS analyses and for systematic comparisons to lattice calculations at finite skewness.\\

\section*{Open data statement}
To promote open and reproducible research, the computer code used to generate the numerical results and figures in this paper is available on GitHub \cite{stringy-gpds} and Zenodo \cite{zenodo15738460}.

\begin{acknowledgments}
We thank Martha Constantinou and Shunzo Kumano for discussions.
F.H.\ is funded by the Austrian Science Fund (FWF) [10.55776/J4854]. F.H.\ thanks Fabio Leimgruber for assistance with code development and troubleshooting, and Kirill M.\ Semenov-Tian-Shansky for pointing out similarities with other approaches.
K.M.\ is supported by DOE Grant No.\ DE-FG02-04ER41302 and by Contract No.\ DE-AC05-06OR23177 under which Jefferson Science Associates operates Jefferson Lab.
I.Z.\ is supported by the U.S.\ Department of Energy under Grant No.\ DE-FG-88ER40388.
This research forms part of the Quark--Gluon Tomography Topical Collaboration, Contract No.\ DE-SC0023646.
\end{acknowledgments}

% ======================================================================
\appendix
\begin{widetext}
\section{Kinematics and rapidity gap}
\label{app:kin}

The kinematics of DVCS allow an interpretation of the skewness $\eta$ in terms of a rapidity gap $\Delta y$.
Let $p_{1}^{\mu}$ and $p_{2}^{\mu}$ be the initial and final nucleon momenta with average $P^{\mu}=\tfrac12(p_{1}^{\mu}+p_{2}^{\mu})$ and transfer $\Delta^{\mu}=p_{2}^{\mu}-p_{1}^{\mu}$.
With $\bm P_{\perp}=0$ one has $p_{1}^{+}=P^{+}(1+\eta)$ and $p_{2}^{+}=P^{+}(1-\eta)$.
For an on-shell four-vector with $\bm p_{\perp}=0$ the longitudinal rapidity is
\begin{equation}
  y=\tfrac12\ln\!\Bigl(\tfrac{p^{+}}{p^{-}}\Bigr)
    =\ln\!\Bigl(\tfrac{p^{+}}{m_{N}}\Bigr),
\end{equation}
where we used $p^{+}p^{-}=m_{N}^{2}$.
The rapidity gap (magnitude) between incoming and outgoing nucleons is
\begin{equation}
  \Delta y=|y_{1}-y_{2}|
          =\ln\!\Bigl(\tfrac{p_{1}^{+}}{p_{2}^{+}}\Bigr)
          =\ln\!\Bigl(\tfrac{1+\eta}{1-\eta}\Bigr),
\end{equation}
or equivalently $\eta=\tanh\!\bigl(\tfrac{\Delta y}{2}\bigr)$, which is Eq.~\eqref{eq:rapidity_gap} in the main text.

\section{Derivation of Eq.~\texorpdfstring{\eqref{eq:NormVariationFinal}}{(Norm)}}
\label{app:SkewNorm}

Starting from the Hankel representation
\begin{equation}
\rho_{n}^{\mathcal H_a}(b_\perp,\eta;\mu)=
   \frac{1}{2\pi}\!\int_{0}^{\infty}\!\dd q\,q\,
   J_{0}(qb_\perp)\,
   \mathcal H_{a}(q^{2}+c_\eta,\eta;\mu),
\end{equation}
the plane integral reads
\begin{equation}
N_{n}^{\mathcal H_a}(\eta;\mu)=
   2\pi\!\int_{0}^{\infty}\!b_\perp \dd b_\perp\,\rho_{n}^{\mathcal H_a}(b_\perp,\eta;\mu).
\end{equation}
Exchanging integrations and using $\int_{0}^{\infty}b\,\dd b\,J_{0}(qb)=\delta(q)/q$ gives
\begin{equation}
N_{n}^{\mathcal H_a}(\eta;\mu)=
   \int_{0}^{\infty}\!\dd q\,q\,\mathcal H_{a}(q^{2}+c_\eta,\eta;\mu)\,\frac{\delta(q)}{q}
   =\mathcal H_{a}(c_\eta,\eta;\mu).
\end{equation}
Because the integrand in Eq.~\eqref{eq:bspaceFT} depends only on $\Delta_\perp=|\Dperp|$, the Fourier integral separates into two cylindrically symmetric Hankel transforms,
\begin{subequations}
\begin{align}
\rho^{\mathcal H_a}_{n}(b_\perp,\eta;\mu)
  &=\frac{1}{2\pi}\!
    \int_{0}^{\infty}\!\dd\Delta_\perp\,\Delta_\perp\,
    J_{0}(\Delta_\perp b_\perp)\,
    \mathcal H_{a}\!\left(n,\eta,t;\mu\right),
\\[2pt]
\rho^{\mathcal E_a}_{n}(b_\perp,\eta;\mu)
  &=\frac{1}{2\pi}\!
    \int_{0}^{\infty}\!\dd\Delta_\perp\,\Delta_\perp^{2}\,
    J_{1}(\Delta_\perp b_\perp)\,
    \mathcal E_{a}\!\left(n,\eta,t;\mu\right),
\end{align}
\end{subequations}
with
\begin{align*}
    \mathcal H_{a}(n,\eta,t;\mu)&\equiv
       \mathbb H_{a}^{(\pm)}\!\left(n,\eta,-\frac{\Delta_\perp^{2}}{1-\eta^2}-c_\eta;\mu\right),\\
    \mathcal E_{a}(n,\eta,t;\mu)&\equiv
       \mathbb E_{a}^{(\pm)}\!\left(n,\eta,-\frac{\Delta_\perp^{2}}{1-\eta^2}-c_\eta;\mu\right).
\end{align*}
Therefore,
\begin{equation}
\rho^{a}_{n}(\bperp;\eta;\mu)
   =\rho^{\mathcal H_a}_{n}(b_\perp,\eta;\mu)
    +\frac{b_{y}}{2m_{N}b_\perp}\,
     \rho^{\mathcal E_a}_{n}(b_\perp,\eta;\mu).
\end{equation}
The full strength follows as
\begin{align}
N^{a}_{n}(\eta;\mu)
   &=\!\int\!\dd^{2}\bperp\,\rho^{a}_{n}(\bperp;\eta;\mu)
    =\!\int\!\dd^{2}\bperp\,\rho^{\mathcal H_a}_{n}(b_\perp,\eta;\mu)
    + \frac{1}{2m_{N}}
     \int\!\dd^{2}\bperp\,
     \frac{b_{y}}{b_\perp}\;
     \rho^{\mathcal E_a}_{n}(b_\perp,\eta;\mu).
\end{align}
The second term vanishes by angular integration, yielding Eq.~\eqref{eq:NormVariationFinal}.
\end{widetext}
% ======================================================================
\clearpage
\bibliography{main}

\end{document}